\documentclass[10pt,twocolumn,letterpaper]{article}

\usepackage{wacv}
\usepackage{times}
\usepackage{epsfig}
\usepackage{graphicx}
\usepackage{amsmath}
\usepackage{amssymb}
\usepackage{hyperref}
\usepackage{subfig}
\usepackage{color,soul}
\usepackage{gensymb}
\usepackage{url}

\wacvfinalcopy % *** Uncomment this line for the final submission

\def\wacvPaperID{****} % *** Enter the wacv Paper ID here

\ifwacvfinal\fi
\begin{document}

%%%%%%%%% TITLE
\title{A Two Stage GAN for High Resolution Retinal Image Generation and Segmentation}

% Authors at the same institution
%\author{First Author \hspace{2cm} Second Author \\
% Authors at different institutions
\author{Paolo Andreini \\
University of Siena\\
\and
Simone Bonechi \\
University of Siena\\
\and
Monica Bianchini \\
University of Siena\\
\and
Alessandro Mecocci \\
University of Siena\\
\and
Franco Scarselli \\
University of Siena\\
\and
Andrea Sodi \\
University of Florence\\
}

\maketitle
\ifwacvfinal\thispagestyle{empty}\fi

%%%%%%%%% ABSTRACT
\begin{abstract}
In recent years, the use of deep learning is becoming increasingly popular in computer vision. However, the effective training of deep architectures usually relies on huge sets of annotated data. 
This is critical in the medical field where it is difficult and expensive to obtain annotated images. In this paper, we use Generative Adversarial Networks (GANs) for synthesizing high quality retinal images, along with the corresponding semantic label--maps, to be used instead of real images during the training process.
Differently from other previous proposals, we suggest a two step approach: first, a progressively growing GAN is trained to generate the semantic label--maps, which describe the blood vessel structure (i.e. vasculature); second, an image--to--image translation approach is used to obtain realistic retinal images from the generated vasculature. By using only a handful of training samples, our approach generates realistic high resolution images, that can be effectively used to enlarge small available datasets. Comparable results have been obtained employing the generated images in place of real data during training. The practical viability of the proposed approach has been demonstrated by applying it on two well established benchmark sets for retinal vessel segmentation, both containing a very small number of training samples. Our method obtained better performances with respect to state--of--the--art techniques.
\end{abstract}

%%%%%%%%% BODY TEXT
\section{Introduction} 

The retinal microvasculature is the only part of the human circulation that can be directly and non--invasively visualized \textit{in vivo} \cite{PATTON200699}. Hence, it can be easily acquired and analyzed by automatic tools.
As a result, retinal fundus images have a multitude of applications, ranging from biometric identification, to computer--assisted laser surgery, to the diagnosis of several disorders \cite{FRAZ2012407}.
One important processing step in such applications is the proper segmentation of the retinal vessels. For this reason, we propose a new deep learning approach for retinal image generation and vessel segmentation.
Image semantic segmentation aims at making dense predictions by inferring the object class for each pixel of an image. The segmentation of digital retina images allows to extract various quantitative vessel parameters, in order to obtain more objective and accurate medical diagnosis. In particular, the segmentation of retinal blood vessels, can help the diagnosis, treatment, and monitoring of diseases such as diabetic retinopathy, hypertension, and arteriosclerosis \cite{bowling2015kanski,abramoff2010retinal}.

\begin{figure}[!ht]
\begin{center}
\centerline{\includegraphics[scale=0.30]{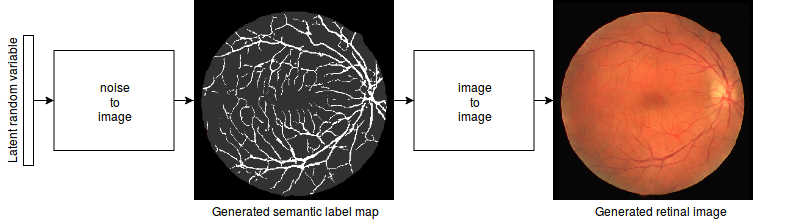}}
\caption{The proposed two--step image generation scheme.}
\label{general_scheme}
\end{center}
\end{figure}

It is widely recognized that Deep Neural Networks (DNNs) are becoming the standard approach in semantic segmentation \cite{FCN,deeplab,PSP} and in many other computer vision tasks \cite{AlexNet,pose,maskrcnn}.
DNN training, however, requires large sets of accurately labeled data, so that the availability of annotated images is becoming increasingly critical. This is particularly true in medical applications where data collection is often difficult and expensive. For this reason, generating synthetic data is of great interest. Nevertheless, synthesizing high resolution realistic medical images remains a complex unsolved challenge.
Most of the leading approaches for semantic segmentation rely on thousands of supervised images while supervised public datasets for retinal vessel segmentation are very small (most datasets contain less than 30 images). To face the scarcity of data, we propose a new approach for the generation of retinal images along with the corresponding semantic label--maps. Specifically, we propose a novel generation procedure based on two distinct phases. In the first phase, a generative adversarial network (GAN) \cite{goodfellow2014generative} learns to generate the blood vessel structure (i.e. the vasculature). The GAN is trained to learn the typical semantic label--map--distribution from a small set of training samples. To generate high resolution label--maps, the Progressively Growing GAN \cite{Karras2018ProgressiveGO} (PGGAN) approach has been employed. In a second, and distinct phase, an image--to--image translation algorithm \cite{Wang2018HighResolutionIS} is used to translate the blood vessels structures into realistic retinal images (see Fig. \ref{general_scheme}).

The rationale behind this approach is that, in many applications, the semantic structure of an image can be learned regardless of its visual appearance. 
Once the semantic label map has been generated, visual details can be incorporated using an image--to--image translation algorithm, thus obtaining realistic synthesized images. Separating the whole process into two phases, we obtained retinal images with an unprecedented high resolution and quality, along with their semantic label--maps. Furthermore, the number of samples required for training is significantly reduced. 
% valutare o si scrive che sono pochi senza confrontarsi oppure se c'è tempo si inserisce confronto %
It is worth noting that the proposed two--step approach reduces the GPU memory requirements w.r.t. a single step method.
The generation of the label--maps through a GAN, allows to synthesize a virtually infinite number of different training samples with different vasculature.

To assess the usefulness and correctness of the proposed approach, the generation procedure has been applied on two public datasets (i.e. DRIVE \cite{Staal2004RidgebasedVS} and CHASE\_DB1 \cite{Fraz2012AnEC}).
The generated data have been used to train a Segmentation Multiscale Attention Network (SMANet) \cite{SMANet_text}.

The SMANet is a deep convolutional neural network, originally developed for text segmentation in outdoor/indoor scenes.
Its architecture is based on the Pyramid Scene Parsing Network \cite{PSP}, which is a popular semantic segmentation network. The SMANet employs a two--level convolutional decoder and a multi--scale attention mechanism to better detect small objects at different scales.
This multi--scale detection capability is fundamental in case of retinal vessel segmentation, because the vessels show different characteristics depending on their diameter and spatial location.
Comparable results have been obtained by training the SMANet on the generated images in place of real data. It is interesting to note that, if the network is pre--trained on the synthesized data and then fine--tuned on real images, the segmentation results obtained on the DRIVE dataset come very close to those obtained by the best state--of--the--art approach \cite{Sekou2019FromPT}.
If the same approach is applied to the CHASE\_DB1 benchmark, the results overcome (to the best of our knowledge) those obtained by any other previously proposed method. 

The paper is organized as follows. In Section \ref{Related}, the related literature is reviewed. Section \ref{Method} presents a description of the proposed approach. Section \ref{Experiments} shows and discusses experimental results. Finally, Section \ref{Conclusions} draws conclusions and future perspectives.

\section{Related works} \label{Related}
\subsection{Synthetic Image Generation} \label{Synthetic_Image_Generation}
Methods for generating images are by no means new, and can be classified into two main categories: model--based approaches and learning--based approaches.
The most conventional approach is to formulate a model of the observed data and to render the images by a dedicated engine. 
This approach has been used, for example, to extend the available datasets of driving scenes in urban environments \cite{game}, \cite{SYNTHIA} or for object detection \cite{hodan2019photorealistic}, and text segmentation \cite{COCO_TS_ICANN}. Also in the field of medical image analysis, synthetic image generation has been extensively employed. For example, realistic digital brain--phantom have been synthesized in \cite{collins1998design} while, more recently, synthetic agar plate images have been generated for image segmentation \cite{andreini2018deep}. The design of specialized engines for data generation requires an accurate model of the scene and a deep knowledge of the specific domain. For this reason, in recent years, the learning--based approach attracted increasing research resources. In this context, machine learning techniques are used to capture the intrinsic spatial variability of a set of training images, so that the specific domain model is acquired implicitly from the data. 
Once the probability distribution that underlies the set of real images has been learned, the system can be used to generate new images that are likely to mimic the original ones.
One of the most successful machine learning model for data generation is the Generative Adversarial Network (GAN) \cite{goodfellow2014generative}. A GAN is composed of two competing networks, a generator $G$ and a discriminator $D$. $G$ is trained to map a latent random variable $ \textbf{z} \in \mathbb{R}^{Z} $ into fake images $\tilde{\textbf{x}}=G(\textbf{z})$, whereas $D$ aims at distinguishing the fake samples 
%$\textbf{G(z)} \sim p_{G}(z) $ preso da Generative Adversarial Learning for Reducing... non mi tornava la notazione per z vedere se ora è ok
 from real ones, $\textbf{x} \in p_{data}(x) $. 
 The GAN training is formulated as a min--max game between $G$ and $D$:
\begin{align*}\scriptstyle
\min_{G}\max_{D}V(D,G)= \mathbb{E}_{\textbf{x} \sim{p_{r}(x)}}[\log{D(\textbf{x})}] + \mathbb{E}_{\textbf{z}\sim{p_{z}(\textbf{z})}}[\log(1-D(G(\textbf{z})))]
\end{align*}

\noindent One example of GANs used for data augmentation in the medical field has been given in \cite{FridAdar2018SyntheticDA} for the classification of liver lesions, and in \cite{Shin2018MedicalIS} to generate synthetic abnormal MRI images containing by brain tumors.

\subsection{Image--To--Image Translation} \label{image2image}

\noindent Recently, beside image generation, adversarial learning has been also extended to the image--to--image translation, whose goal is to translate an input image from one domain to another.
Many computer vision tasks, such as image super--resolution \cite{ledig2017photo}, image inpainting \cite{Pathak2016ContextEF}, and style transfer \cite{gatys2015neural} can be casted into the image--to--image translation framework.
Both unsupervised \cite{Liu2017UnsupervisedIT}, \cite{NIPS2016_6544}, \cite{Yi2017DualGANUD}, \cite{Zhu2017UnpairedIT} and supervised approaches can be used \cite{Isola2017ImagetoImageTW}, \cite{Karras2018ProgressiveGO}, \cite{Chen2017PhotographicIS}.
Supervised training uses a set of pairs of corresponding images $\left \{(s_i,t_i)\right \}$, where $s_i$ is an image of the source domain and $t_i$ is a corresponding image in the target domain. As an example, Pix2Pix \cite{Isola2017ImagetoImageTW} consists of a conditional GAN that operates in a supervised way, and Pix2PixHD \cite{Karras2018ProgressiveGO} employs a coarse--to--fine generator and discriminator along with a feature--matching loss--function, to translate images with higher resolution and quality.

\subsection{Retinal Image Synthesis} \label{Retinal image synthesis}
One of the first applications of retinal image synthesis has been described in the seminal work \cite{Sagar1994AVE}, in which an anatomic model of the eye and of the surrounding face has been implemented for surgical simulations. More recently, in \cite{stag.20141238}, a large dictionary of small image patches containing no vessels, has been used to model the retinal background and fovea. A parametric intensity model, whose parameters have been estimated from real images, is used to generate the optical disk.
Complementary to \cite{stag.20141238}, the contribution in \cite{10.1007/978-3-319-46630-9_17} focuses on the generation of the vascular network, based on a parametric model whose parameters are learned from real vessel trees. Despite these methods provide reasonable results, they are complex and heavily depend on domain knowledge. To reduce the domain knowledge requirements, a completely learning--based approach has been proposed in \cite{Towards}, where an image--to--image translation model has been employed to transform existing vessel networks into realistic retinal images. The vessel networks used for learning have been obtained using a suitable segmentation technique applied to a set of real retina images. However, the quality of the generated images heavily depends on the performances of the segmentation module. In \cite{ZHAO201814}, a generative adversarial approach, together with a style transfer algorithm, is used to reduce the need for annotated samples and to improve the representativeness (e.g. variability) of the synthesized images. The model still relies on pre--existing vessel--networks (obtained manually or by a suitable segmentation technique). In \cite{end-to-end}, the use of an adversarial auto--encoder for the task of retinal--vessel synthesis has been adopted to avoid the dependence of the model on the availability of preexistent vessel maps. Nevertheless, this approach allows to generate only low resolution images, and the performance in vessel segmentation by using the synthesized data, is far below the state--of--the--art. Higher resolution retinal images, along with their segmentation label--maps, have been generated in \cite{Beers2018HighresolutionMI} with an approach based on a Progressively Growing GAN (PGGAN)\cite{Karras2018ProgressiveGO}. The method allows to generate images up to a resolution of $512\times{512}$ pixels. A set of 5550 images, segmented by a pre--trained U--Net \cite{Ronneberger2015UNetCN} have been used during training. Unfortunately, the usefulness of the generation for image segmentation is not demonstrated.
% forse è da rimarcare che loro fanno 1 step e il risultato fa schifo...
\noindent The present paper improves previous approaches generating synthetic images up to a resolution of $1024\times{1024}$ pixels. The generation is based on a very small set of preexisting images (actually, 20 images with supervised segmentation maps). Both the retinal images and the corresponding semantic label--maps (the vasculature) are generated. Furthermore, we prove that combining real retinal images with synthesized ones during the training of a segmentation network, improves the final segmentation performance.

\subsection{Retinal Vessel Segmentation} \label{Blood Vessels segmentation}
During the last decades, several approaches for retinal vessel segmentation have been proposed, both supervised and unsupervised. Unsupervised methods depend heavily on prior knowledge about the vessel structure.
For example, Vessel Tracking Techniques define an initial set of seed points and, thereafter, by chaining pixels that minimize a given cost function, the vasculature is iteratively extracted \cite{Recursive1993}\cite{YIN20121235}.
In \cite{845178}, retinal images are convolved with a 2D filter to produce a Gaussian intensity profile of the blood vessels, that are subsequently thresholded to give the vessels map. Adaptive thresholding has been used in \cite{Roychowdhury} and \cite{CAMARANETO2017182}. An Active Contour Model, that combines intensity and local phase information, is used in \cite{Zhao}. 
Supervised methods are currently the leading techniques in semantic segmentation. In this framework, truth annotations are used to train a classifier aimed at distinguishing the vessels from the background. Various classification models have been employed for blood--vessel segmentation, based on a preliminary feature engineering stage.
A k--Nearest Neighbor classifier is used in \cite{Meindert}, which adopts a pixel--wise feature vector, based on Gaussian functions and their derivatives.
In \cite{Soares}, a Gaussian Mixture Model classifier is applied to the pixel intensities augmented by coefficients obtained through a 2D--Gabor Wavelet Transform, evaluated at multiple scales.
In \cite{Marin}, a neural network is used to classify vectors comprising gray--levels and moment invariant features.
A random forest classifier, applied on a 29--dimensional feature vector, has been used in \cite{ZHANG2017107}. 
Supervised methods are strongly affected by the feature engineering stage.

Deep learning--based methods automatically learn from the input data an increasingly complex hierarchy of features, bypassing the need for problem specific knowledge. 
In retinal image segmentation a deep convolutional neural network (DCNN) in used in \cite{11} where the training examples are subject to various preprocessing and augmented based on geometric transformations and gamma corrections.
A neural network that can be efficiently used in real--time on embedded systems is proposed in \cite{23}.
In \cite{25}, it is employed a fully convolutional network \cite{FCN} with an AlexNet \cite{AlexNet} encoder. Fully convolutional networks have been used also in \cite{26} and \cite{27}. In \cite{12} the task of segmentation is remolded into a problem of cross--modality data transformation from retinal images to vessel map.
A modified U--Net \cite{Ronneberger2015UNetCN} is used in \cite{Yan2018JointSA} which exploits a combination between a segment--level loss and a pixel--level loss, to deal with the unbalanced ratio between thick and thin vessels in fundus images. 
A Holistically--Nested Edge Detection (HED) network \cite{Xie}, originally designed for edge detection, followed by a conditional random field are employed for the retinal blood vessel segmentation in \cite{30}. Deep supervision is incorporated in some intermediate layers of a VGG network \cite{VGG} in \cite{33} and \cite{34}. In \cite{OLIVEIRA2018229} a Fully Convolutional Neural Network uses a stationary wavelet transform pre--processing step to improve the network performance. Finally, in \cite{Sekou2019FromPT}, a CNN is pre--train on image patches and then fine tuned at the image level.

\section{Retinal Image Generation} \label{Method}
The main goal of this work is to generate realistic retinal images and the corresponding semantic segmentation masks, by using a very small number of training samples. 
The proposed generation procedure is composed of two different phases: the first one is related to the generation of semantic label-maps of the vessels, and the second to the synthesis of realistic images starting from those label-maps.
The quality and usefulness of the generated images have been validated by the performance obtained on two public benchmark datasets using the synthesized images to train a segmentation network.
In particular, Section \ref{Generating the Vasculature} gives an overview of the approach used to generate the semantic label--maps.
Section \ref{Translating Vessels to Retinal Images} describes the image--to--image translation algorithm which synthesizes retinal images from the semantic label--maps. Instead, Section \ref{smanet} describes the semantic segmentation network used to segment the retinal vessels.
Finally, some details about the training method are reported in Section \ref{training_details}.

% valutare se sintetizzare e aggiungere dettagli sulle modifiche fatte, se c'è spazio...
\subsection{Vasculature Generation} \label{Generating the Vasculature}
\begin{figure}[!ht]
\begin{center}
\centerline{\includegraphics[scale=0.33]{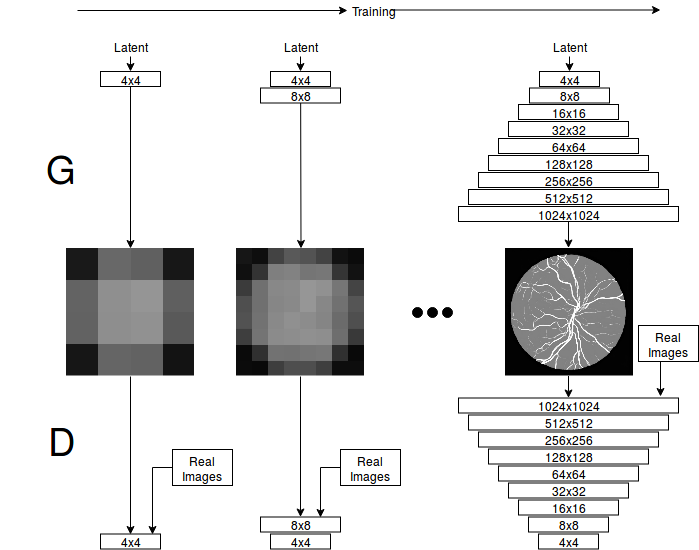}}
\caption{Training schema of the semantic label--maps generation.}
\label{PG_gan_train}
\end{center}
\end{figure}

The generation of the vessel structure is based on the PGGAN approach, capable of learning the distribution of the semantic label--maps. The label--maps are processed to encode both the retinal fundus and the vasculature (i.e. the vessel distribution). 
To reduce the risks related to the lack of an adequate descriptive power, due to the very limited number of available training samples, data augmentation has been applied. Specifically, the semantic label--maps have been slightly rotated ($\pm$ 15\degree) and flipped in different ways (horizontal, vertical and horizontal followed by vertical flips).
The generation starts at low--resolution and then the resolution is progressively increased by adding new layers to the networks. The generator and the discriminator are symmetric and grow in sync. The transition from low--resolution image generation to high--resolution image generation follows the procedure described in \cite{Karras2018ProgressiveGO} to avoid the problems related to a sudden transition.
The training starts with both the generator and the discriminator having a low spatial resolution (e.g. $4\times{4}$ pixels), then the resolution increases progressively until the final resolving power is reached. 
The Wasserstein loss, using a gradient penalty \cite{wgan-gp}, has been used as loss function for the discriminator.
The learning procedure is illustrated in Fig. \ref{PG_gan_train}. 

\noindent  It can be observed that the global structure of the vessel distribution is learned at the beginning of the training, whereas finer details are added as the resolution increases. 
The generation procedure allows to obtain a virtually infinite number of different vasculatures. 
To reduce the probability of introducing artifacts, a simple post--processing has been carried out. Specifically, a morphological opening \cite{morphology} has been applied to the generated retinal fundus mask to improve its circularity. Small holes have been filled, and segments of small dimension have been removed from the generated vessel structure.

\subsection{Translating Vessel Maps into Retinal Images} \label{Translating Vessels to Retinal Images}
Once the vessel networks have been obtained, they must be transformed into realistic color retinal images. 
Our method is based on Pix2PixHD \cite{Karras2018ProgressiveGO}, a supervised image--to--image translation framework based on Pix2Pix \cite{Isola2017ImagetoImageTW}. In Pix2Pix, a conditional GAN learns to generate the output conditioned on the corresponding input image. The generator has an encoder--decoder structure, and takes as input the images belonging to a certain domain $A$ and generates images in a different domain $B$. The discriminator observes couples of images, the image from $A$ is provided as input along with the corresponding image of $B$ (real or generated). The discriminator aims at distinguishing between real and fake (generated) couples. Pix2PixHD improves upon Pix2Pix by introducing a coarse--to--fine generator composed of two subnetworks that operate at different resolutions. A multiscale discriminator is also employed, with an adversarial loss which incorporates a feature--matching loss for training stabilization. In our setup, the semantic label--maps, generated in the previous step, are given in input to the generator that is trained to generate realistic retinal images. Images have been resized to the nearest power--of--two resolution (i.e. the retinal images in the DRIVE dataset that have a resolution of $565\times{584}$ pixels have been resized to $512\times{512}$ pixels, whereas the CHASE dataset images that have a resolution of $999\times{960}$ pixels have been resized to $1024\times{1024}$ pixels).

\begin{figure}[!ht]
\vskip 0.2in
\begin{center}
\centerline{\includegraphics[scale=0.50]{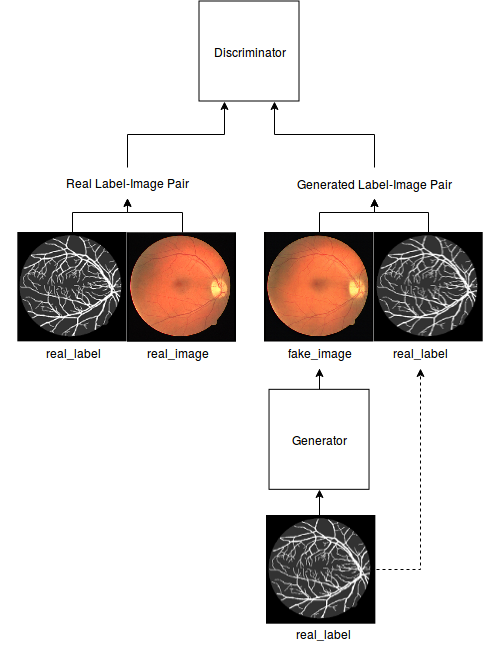}}
\caption{Scheme of the Pix2Pix training framework employed to translate label--maps into retinal images.}
\label{label2retina}
\end{center}
\end{figure}

\noindent An overview of the proposed setup is given in Fig. \ref{label2retina}. 

\subsection{SMANet Architecture}\label{smanet}
\begin{figure*}[!ht]
\vskip 0.2in
\begin{center}
\centerline{\includegraphics[scale=0.40]{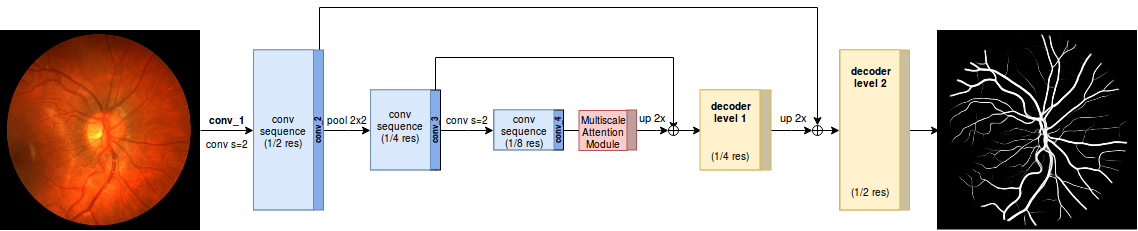}}
\caption{Scheme of the SMANet segmentation network.}
\label{SMANet_arch}
\end{center}
\end{figure*}

The semantic segmentation network employed in this paper is a Segmentation Multiscale Attention Network (SMANet) \cite{SMANet_text}. 
The SMANet, originally proposed for scene text segmentation, comprises three main components: a ResNet encoder, a multi--scale attention module, and a convolutional decoder (see Fig. \ref{SMANet_arch}).

The architecture is based on the PSPNet \cite{PSP}, a deep fully convolutional neural network with a ResNet \cite{ResNet} encoder. In the PSPNet, to enlarge the receptive field of the neural network, a set of standard convolutions of the ResNet backbone has been replaced with dilated convolutions (i.e. atrous convolution \cite{atrous}). Moreover, in the PSPNet, a pyramid of pooling with different kernel size has been employed to gather context information.
The pooled feature maps are then up--sampled at the same resolution of the ResNet output, concatenated and fed into a convolutional layer, to obtain an encoded representation.
In the original PSPNet, this representation is followed by a final convolutional layer that reduces the feature maps to the number of classes. The desired per--pixel prediction, is obtained directly up--sampling to the original image resolution.
In the SMANet, a multi scale attention mechanism is adopted to focus on the relevant objects present in the image while, a two level convolutional decoder is added to the architecture to better handle the presence of thin objects.

\subsection{Training Details}\label{training_details}
The SMANet, used in this work, is implemented in TensorFlow. Random crops of $281\times{281}$ pixels have been employed during training, whereas a sliding window of the same size has been used for the evaluation.
%% verificare se si è usato multiscala e inserirlo oltre a dettagli sulla generazione
The Adam optimizer \cite{kingma2014adam}, based on a learning rate of 10\textsuperscript{-4} and a mini–batch of 17 examples, has been used to train the SMANet.
All the experiments have been carried out in a Linux environment on a single NVIDIA Tesla V100 SXM2 with 32 GB RAM.

\section{Experiments and Results} \label{Experiments}

\subsection{The benchmark datasets} \label{Datasets}
\begin{itemize}
\item \textbf{DRIVE dataset --}
The DRIVE dataset \cite{Staal2004RidgebasedVS} includes 40 retinal--fundus images of size $584 \times 565  \times 3$ (20 images are for training and 20 for test). The images have been collected by a screening program for diabetic retinopathy in the Netherlands.
Among the 40 photographs, 33 show no diabetic retinopathy, while 7 show mild early diabetic retinopathy. Segmentation ground--truth is provided both for training and test sets. 

\item \textbf{CHASE\_DB1 dataset --}
The CHASE\_DB1 dataset \cite{Fraz2012AnEC} is composed by 28 fundus images of size $960 \times 999 \times 3$ corresponding to the left and right eyes of 14 children. Each image is annotated by two independent human experts. An officially defined split between training and test is not provided for this dataset.
In our experiments we have adopted the same strategy of \cite{Yan2018JointSA} and \cite{12}, selecting the first 20 images for training and the remaining 8 for test.
\end{itemize}

\subsection{Experimental Results}\label{Exp_res}
\begin{figure*}[!ht]
\centering
\includegraphics[width=3cm,height=3cm]{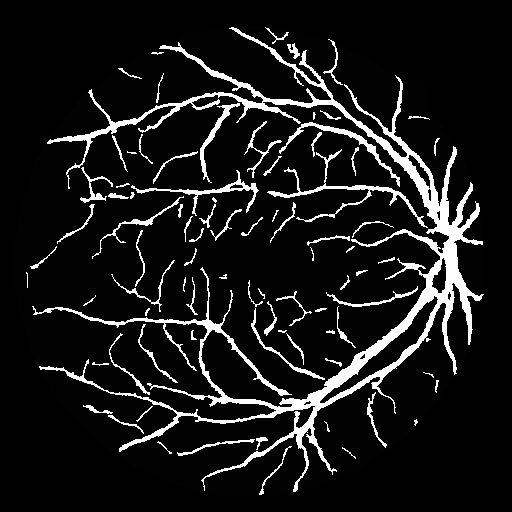}
\includegraphics[width=3cm,height=3cm]{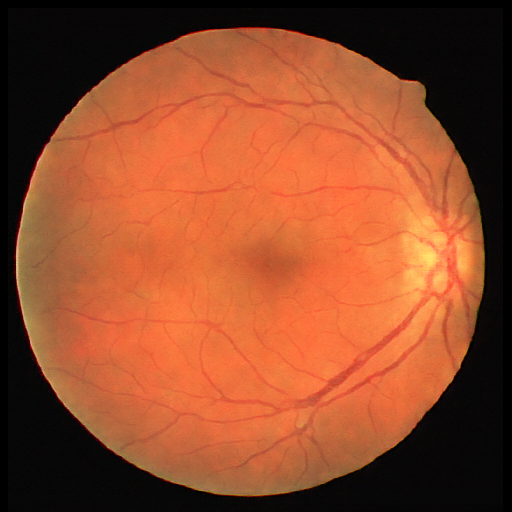}
\hskip 0.3cm
\includegraphics[width=3cm,height=3cm]{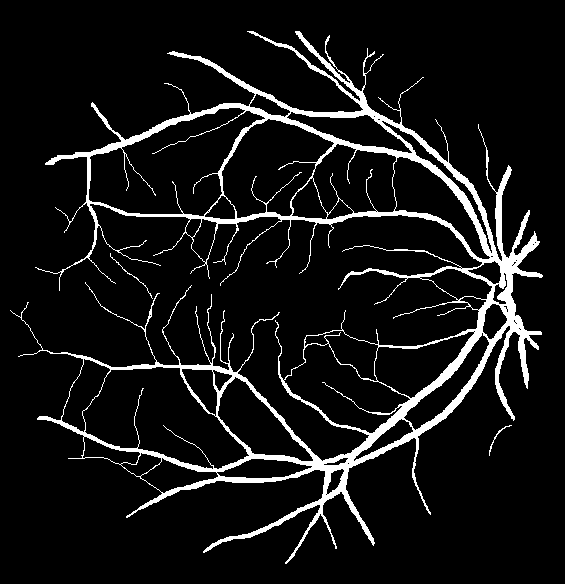}
\includegraphics[width=3cm,height=3cm]{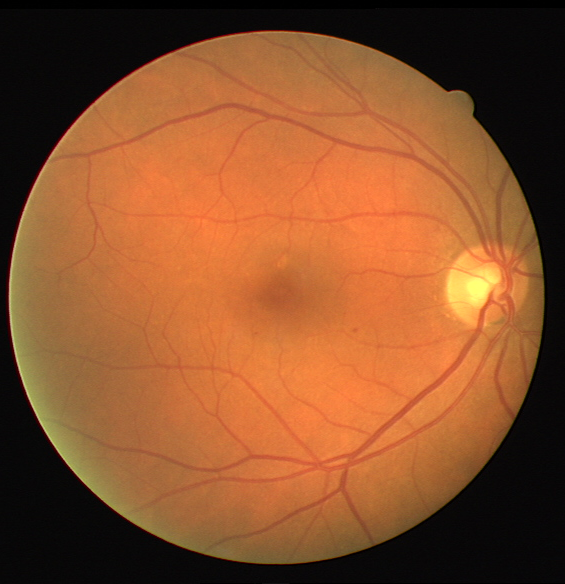}
\\
\includegraphics[width=3cm,height=3cm]{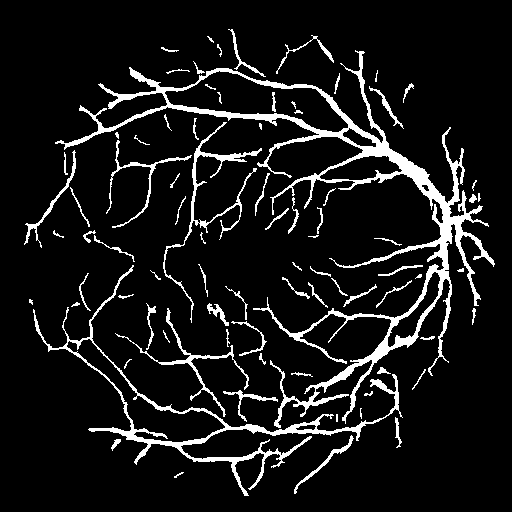}
\includegraphics[width=3cm,height=3cm]{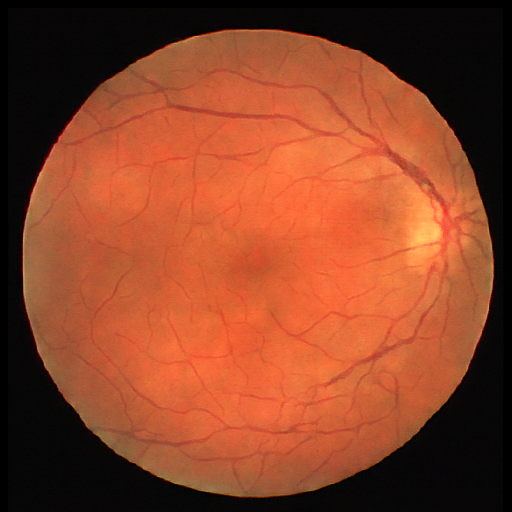}
\hskip 0.3cm
\includegraphics[width=3cm,height=3cm]{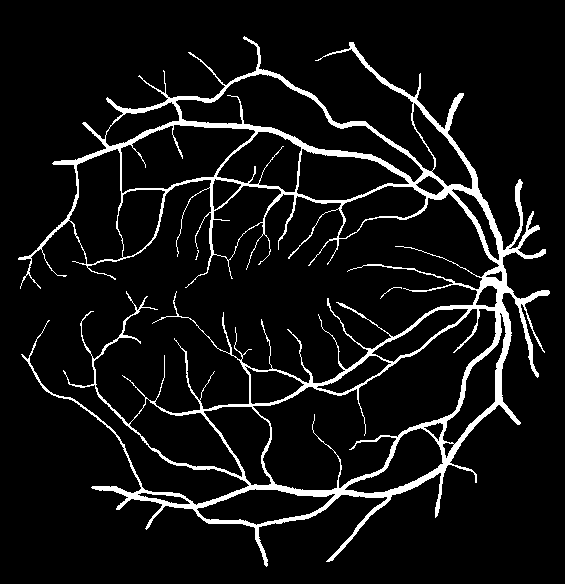}
\includegraphics[width=3cm,height=3cm]{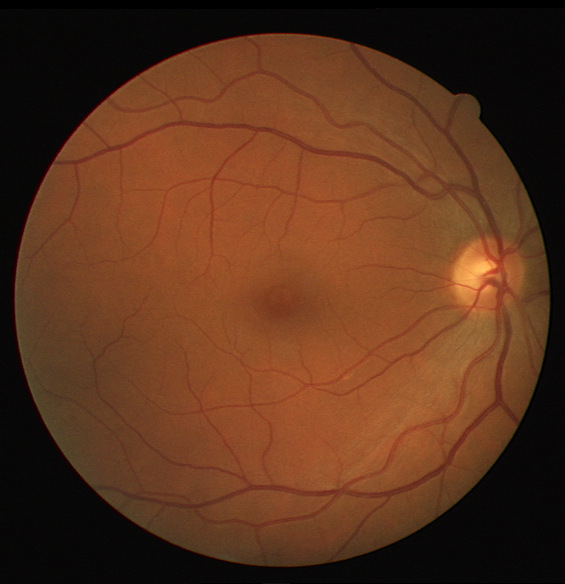}
\caption{Examples of generated (left) and corresponding real images (right) for DRIVE.}
\label{DRIVE_examples}
\end{figure*}

\begin{figure*}[!ht]
\centering
\includegraphics[width=3cm,height=3cm]{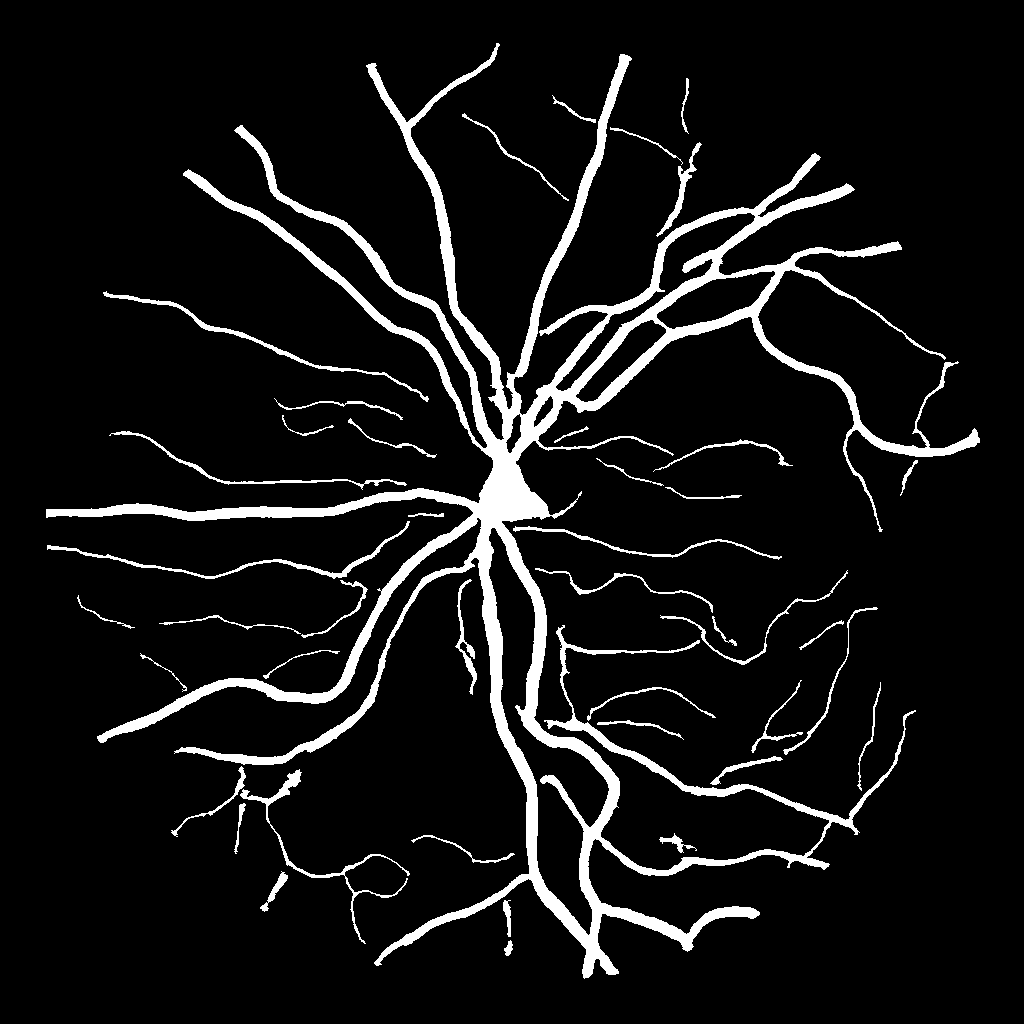}
\includegraphics[width=3cm,height=3cm]{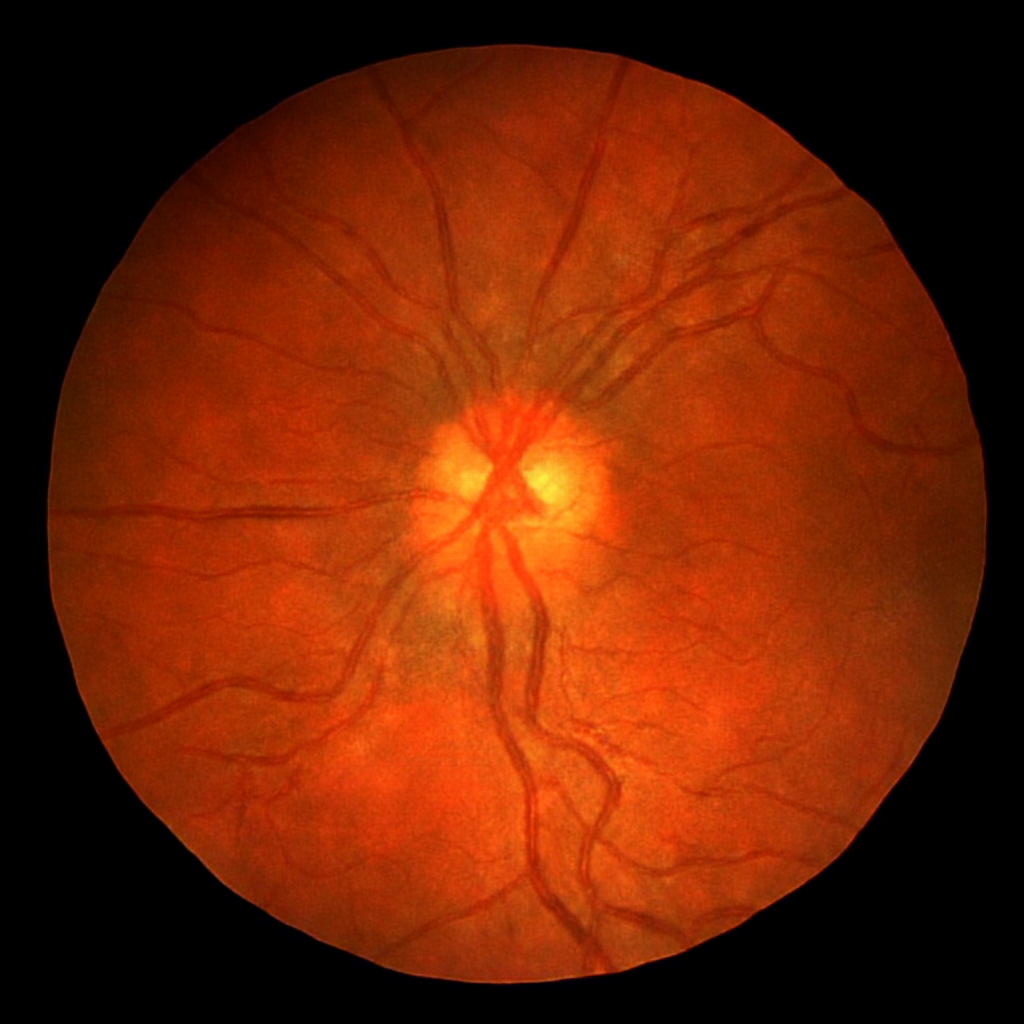}
\hskip 0.3cm
\includegraphics[width=3cm,height=3cm]{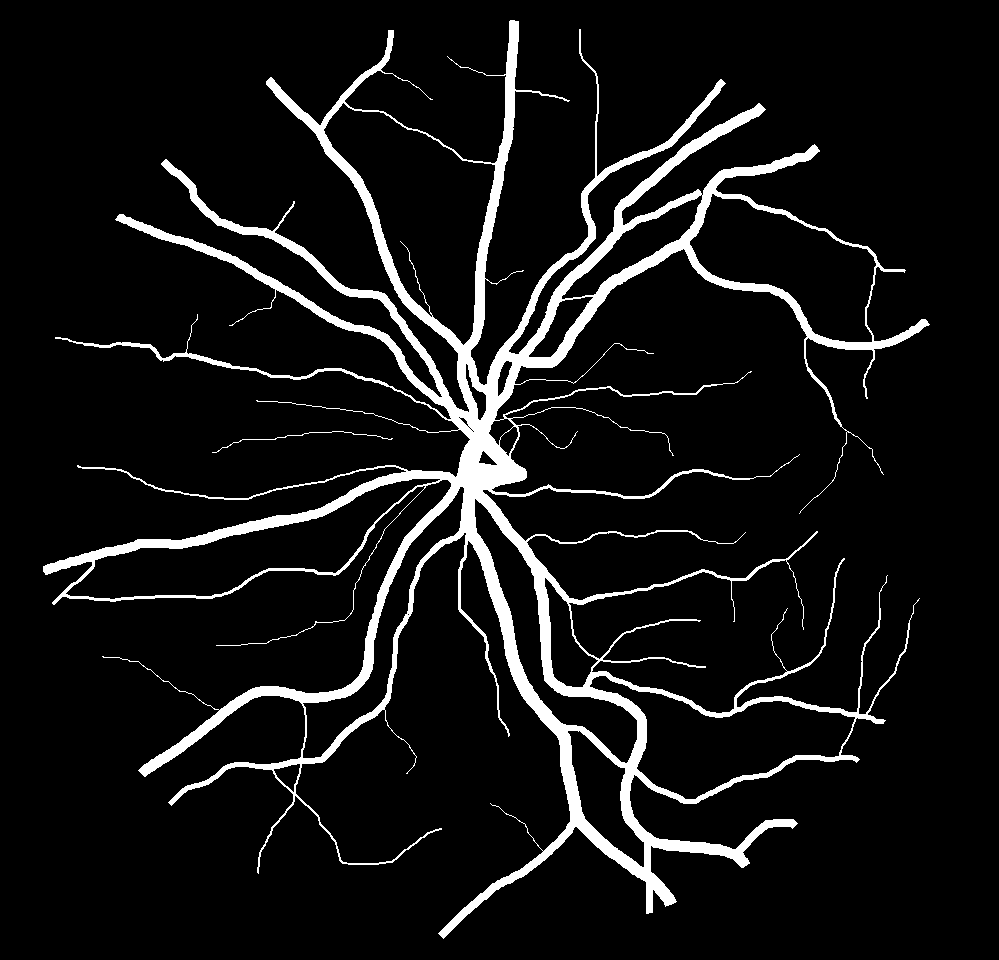}
\includegraphics[width=3cm,height=3cm]{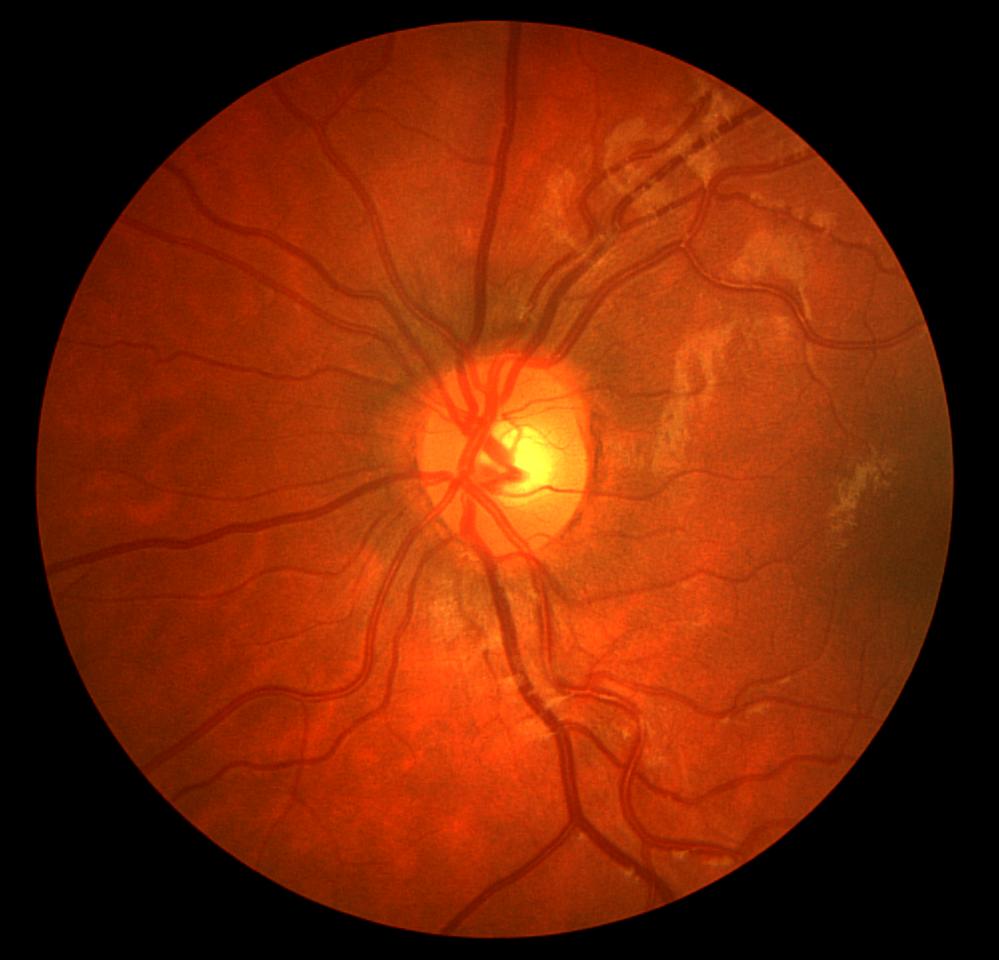}
\\
\includegraphics[width=3cm,height=3cm]{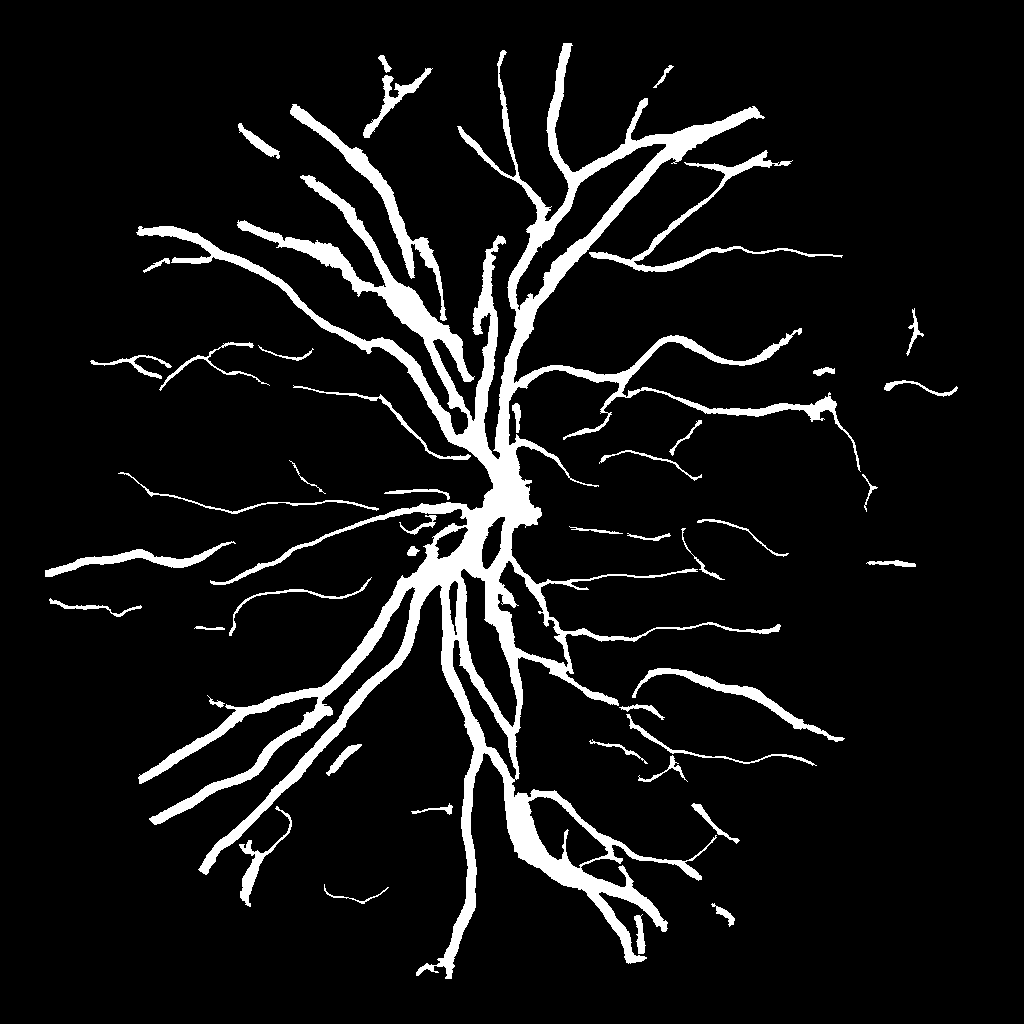}
\includegraphics[width=3cm,height=3cm]{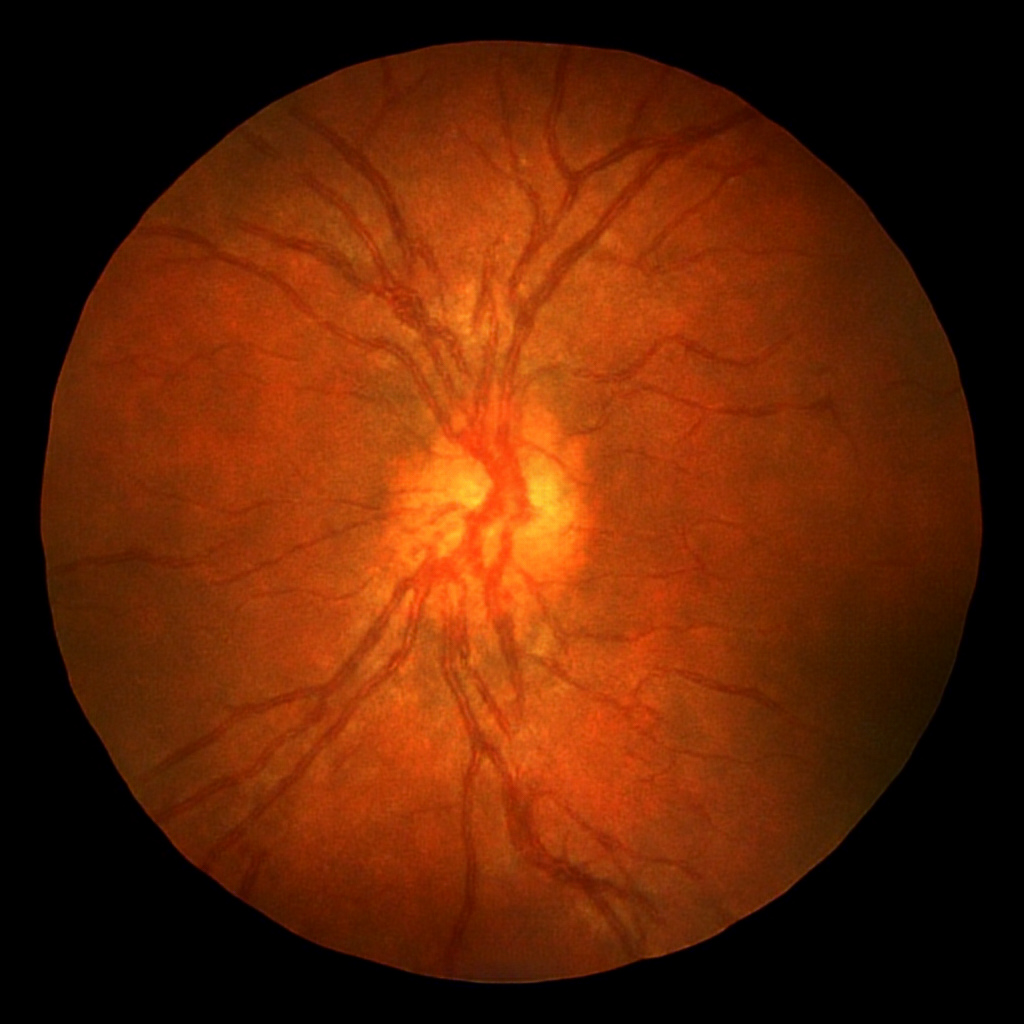}
\hskip 0.3cm
\includegraphics[width=3cm,height=3cm]{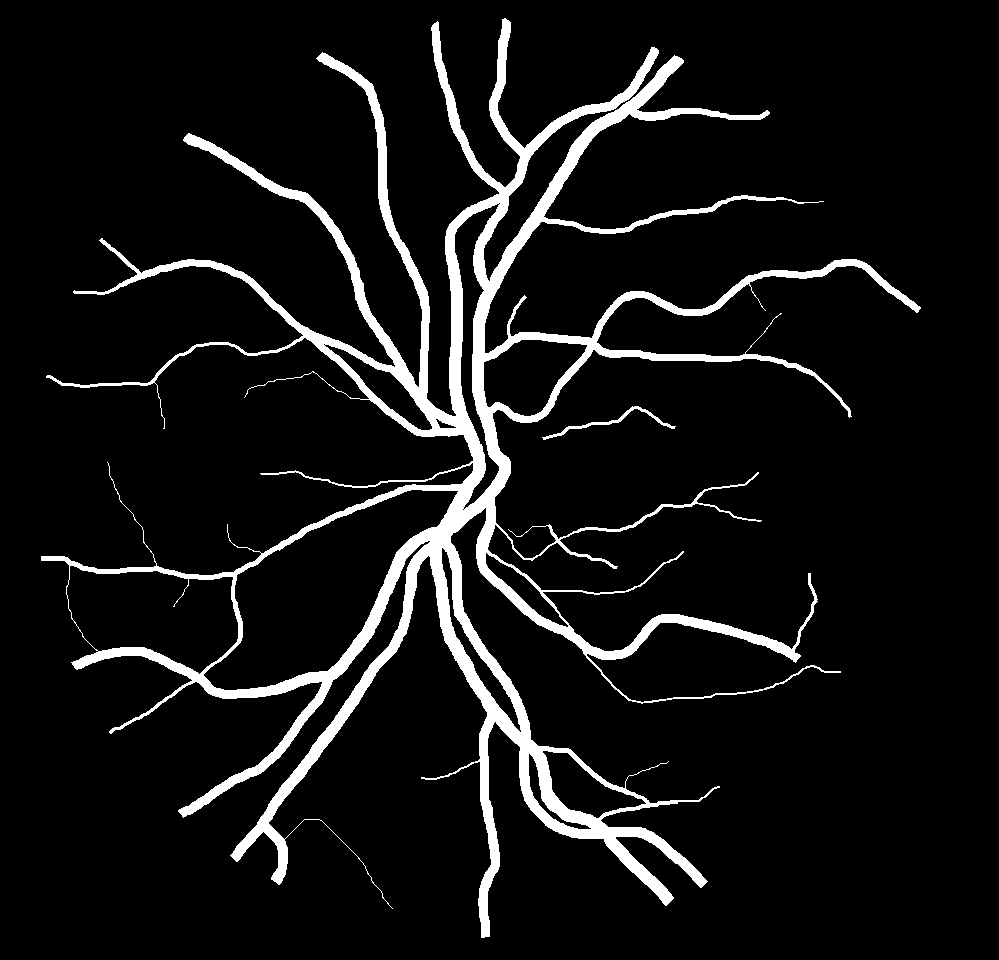}
\includegraphics[width=3cm,height=3cm]{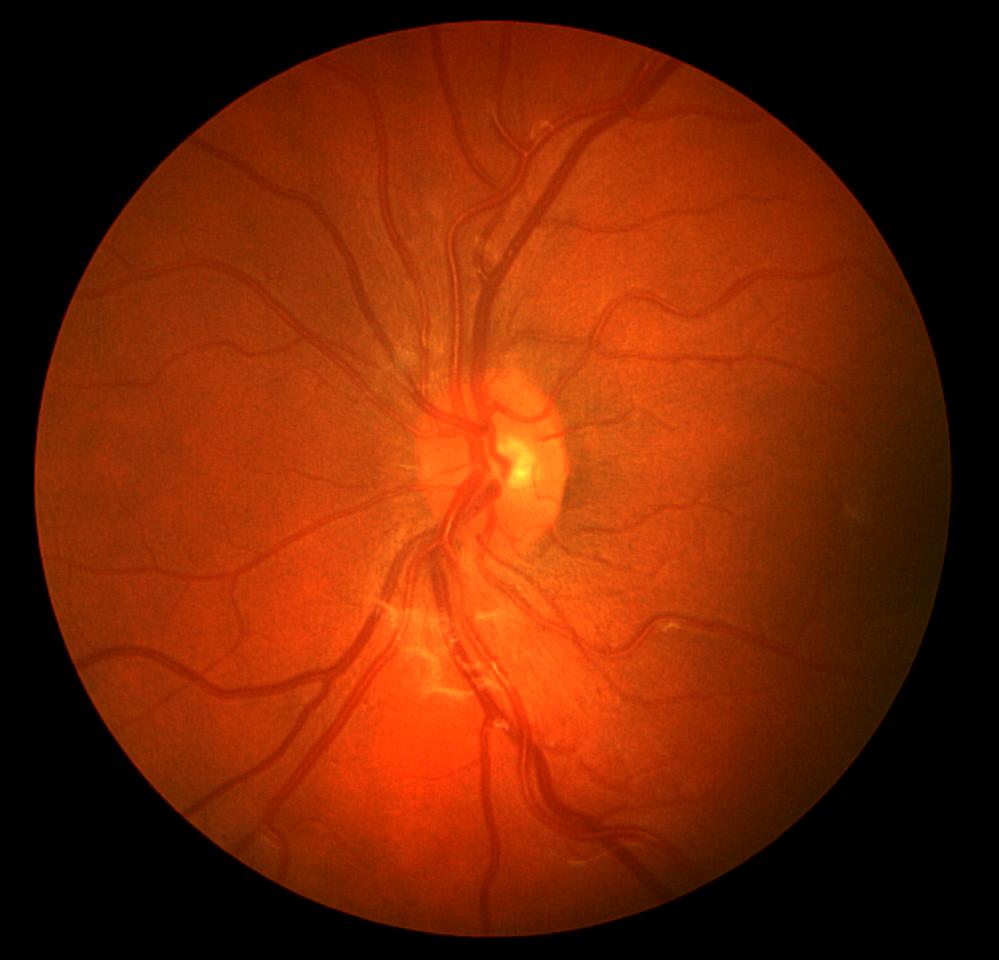}
\caption{Examples of generated (left) and corresponding real images (right) for CHASE\_DB1.}
\label{CHASE_examples}
\end{figure*}

In this paper, we provide both a qualitative and a quantitative evaluation of the generated data. In particular, the quantitative analysis consists in evaluating the usefulness of the generated images for training a semantic segmentation network. This approach is similar to \cite{Shmelkov2018HowGI} and it is based on the assumption that the performances of a deep learning architecture can be directly related with the quality and variety of GAN generated images.
Some qualitative results of the generated retinal images for the DRIVE and CHASE\_DB1 dataset are given in Figs. \ref{DRIVE_examples}--\ref{CHASE_examples}.

\begin{figure*}[!ht]
\vskip 0.2in
\begin{center}
\includegraphics[width=11cm,keepaspectratio]{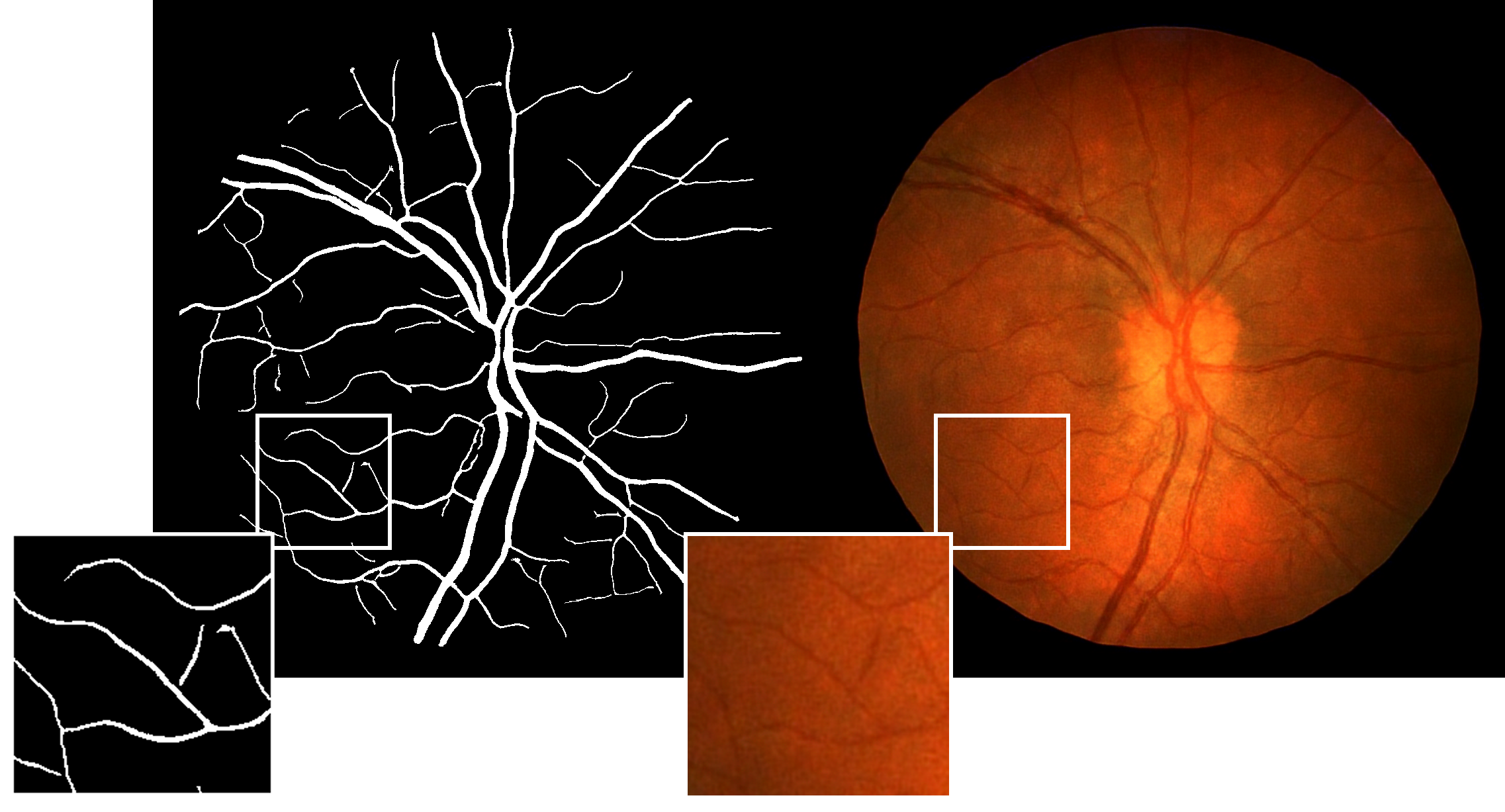}
\caption{An example of a generated image with resolution $1024\times{1024}$, and of the corresponding label map, for the CHASE\_DB1 dataset.}
\label{zoomed}
\end{center}
\end{figure*}

\noindent In Fig. \ref{zoomed}, a zoom on a random patch of a high resolution generated image shows that the image--to--image translation allows to effectively translate the generated vessel structures in retinal images maintaining the semantic information provided by the semantic label map.

\begin{figure*}[!ht]
\vskip 0.2in
\begin{center}
\includegraphics[width=3cm,height=3cm]{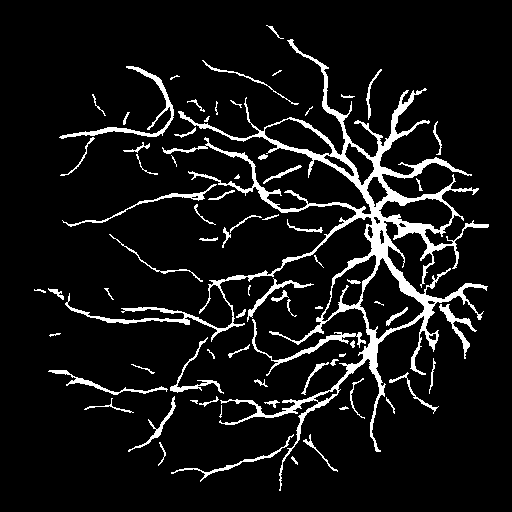}
\includegraphics[width=3cm,height=3cm]{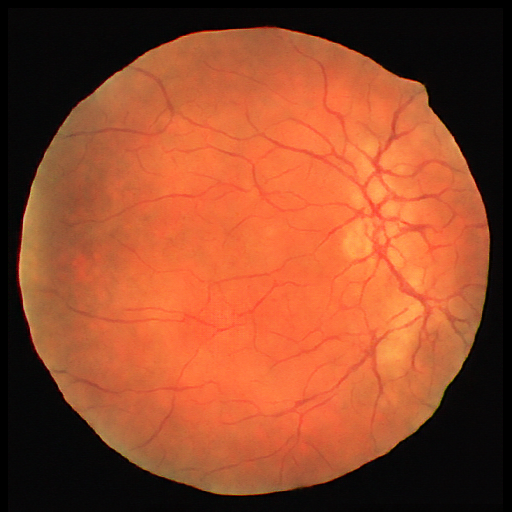}\hskip 0.3cm
\includegraphics[width=3cm,height=3cm]{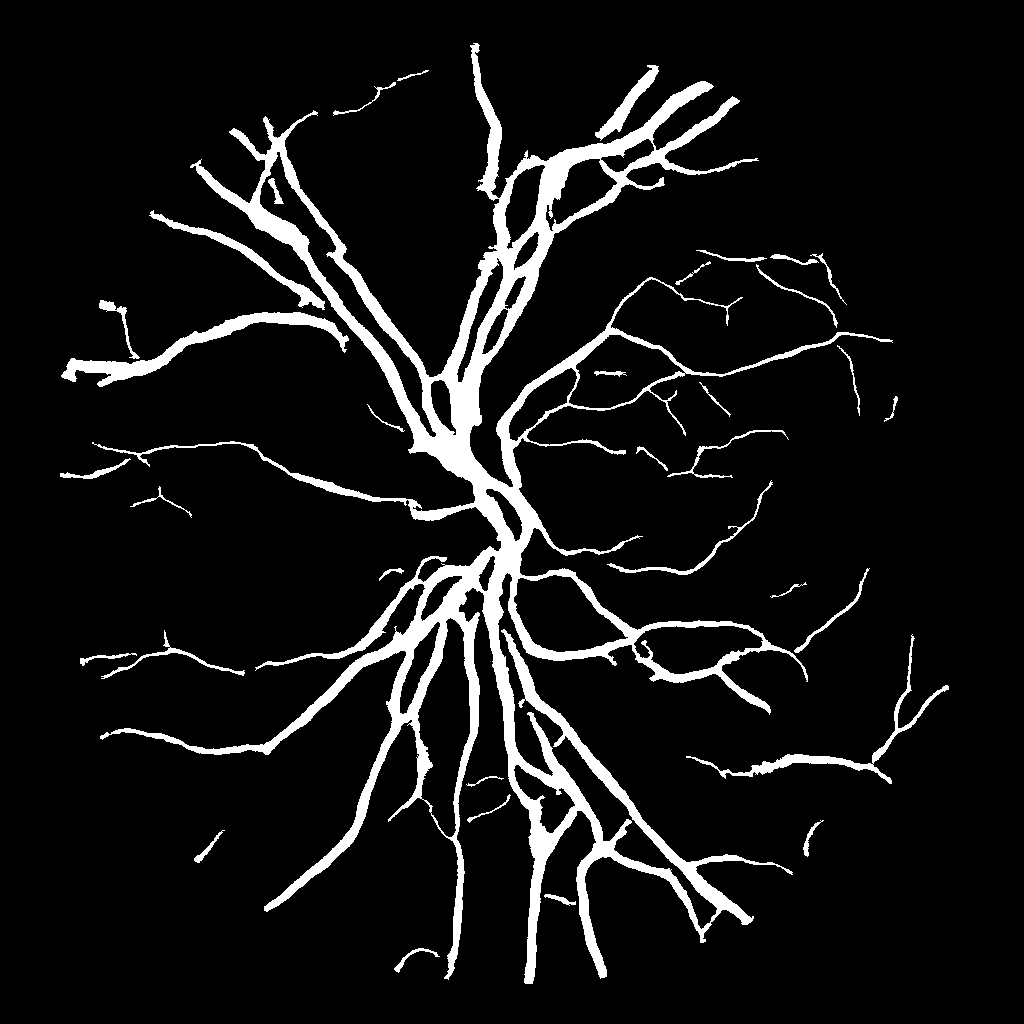}
\includegraphics[width=3cm,height=3cm]{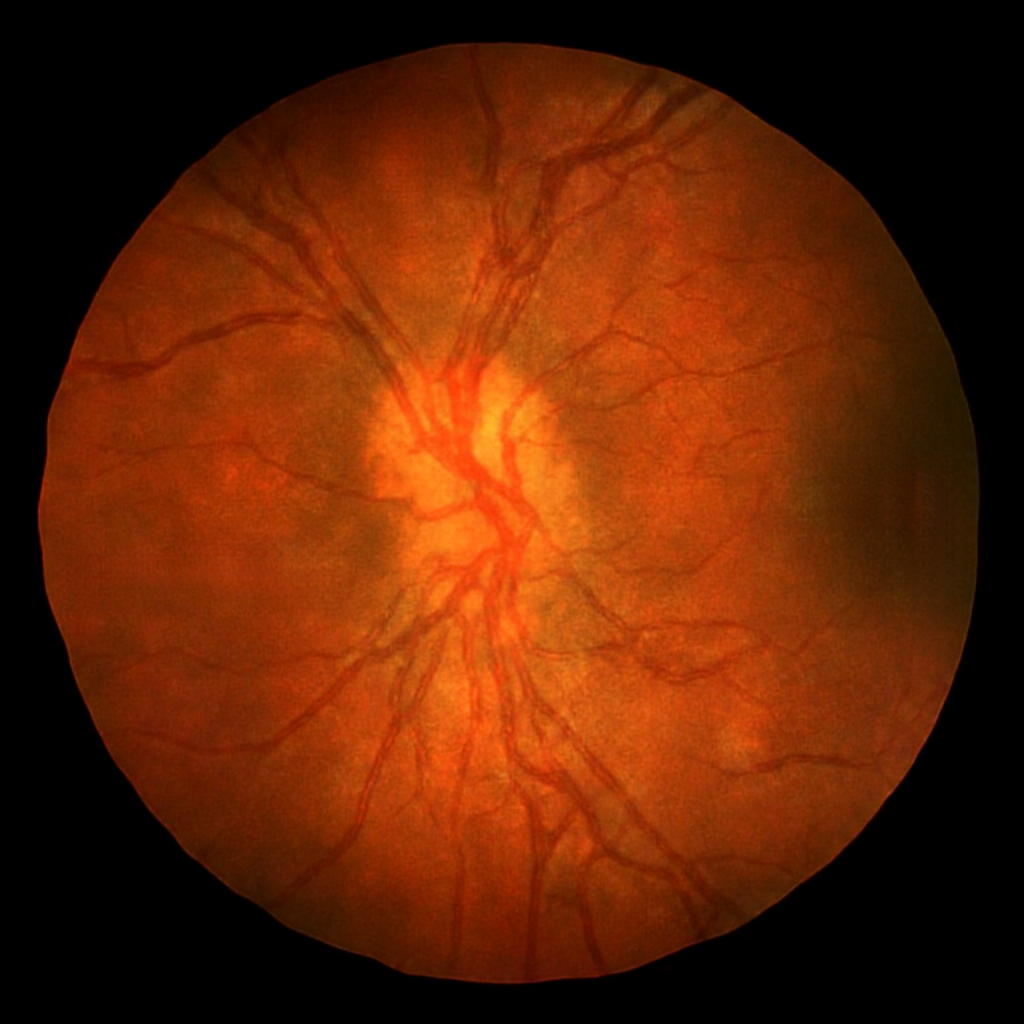}
\caption{Examples of DRIVE and CHASE generated images with unrealistic optical disc and vasculature (left and right respectively).}
\label{artifact}
\end{center}
\end{figure*}

\noindent It must be noted that, even if the major part of the generated samples accurately resembles real retinal fundus images, few examples are evidently suboptimal (see Fig. \ref{artifact} that shows disconnected vessels and an unrealistic optical disc).  

In Table \ref{Generation_comparison}, the proposed method for retinal image generation is compared with other learning--based approaches found in literature.

\begin{table}[ht]
\centering
\fontsize{9}{11}\selectfont  
\begin{tabular}{|l|l|l|l|}
\hline
Methods & Gen. Vessels & Max Res. & Samples\\ \hline
Costa et al. \cite{Towards} & No & $512\times{512}$ & 614\\ 
Zhao et al. \cite{ZHAO201814} & No & $2048\times{2048}$ & 10-20\\ 
Costa et al. \cite{end-to-end} & Yes & $256\times{256}$ & 634\\ 
Beers et al. \cite{Beers2018HighresolutionMI} & Yes & $512\times{512}$ & 5550\\ 
Our & Yes & $1024\times{1024}$ & 20\\
\hline
\end{tabular}
\caption{Comparison among different generation approaches.}
\label{Generation_comparison}
\end{table}

\hl{}

The generation procedure described in Section \ref{Method} has been employed to generate 10000 synthetic retinal images for both the DRIVE and the CHASE\_DB1 datasets. To evaluate the usefulness of the generated data for semantic segmentation, we employed the following experimental set up:
 \begin{itemize}
\item SYNTH -- the semantic segmentation network is trained by using only the 10000 generated synthetic images. 
\item REAL -- only real data are used to train the semantic segmentation network.
\item SYNTH + REAL -- synthetic data are used to pre--train the semantic segmentation network and real data are employed for fine--tuning.
\end{itemize}

\noindent In Table \ref{DRIVE_synth_comparison} and Table \ref{CHASE_synth_comparison}, the results of the vessel segmentation for the DRIVE and CHASE\_DB1 datasets, obtained using the previously described approaches, are respectively reported.

\begin{table}[!ht]
    \centering
    \fontsize{9}{11}\selectfont  
    \begin{tabular}{|l|l|l|l|l|}
    \hline
     Methods & AUC  & Acc\\ \hline
    SYNTH & 98.5  \% & \textbf{97.9} \%\\
    REAL & 98.48 \% & 96.87 \%\\
    SYNTH + REAL & \textbf{98.65} \% & 96.9 \%\\
    \hline
    \end{tabular}
\vspace*{0.3cm}
\caption{\small{Evaluation of the use of generated data on the DRIVE dataset.}}
\label{DRIVE_synth_comparison}
\end{table}

\begin{table}[ht!]
    \centering
    \fontsize{9}{11}\selectfont  
    \begin{tabular}{|l|l|l|l|l|}
    \hline
     Methods & AUC & Acc\\ \hline
    SYNTH & 98.64 \% & 97.49 \%\\
    REAL & 98.82 \% & 97.5 \%\\
    SYNTH + REAL & \textbf{99.16} \% & \textbf{97.72} \%\\
    \hline
    \end{tabular}
\vspace*{0.3cm}
\caption{\small{Evaluation of the use of generated data on the CHASE\_DB1 dataset.}}
\label{CHASE_synth_comparison}
\end{table}

\noindent It can be observed that the semantic segmentation network, trained on synthetic data, produces results very similar to those obtained by training on real data. This demonstrates that the generated images effectively capture the training image distribution, so that they can be used to adequately train a deep neural network. Moreover, if fine--tuning with real data is applied after a pre--training with synthetic data only, the results further improve w.r.t. the use of real data only. This fact indicates that the generated data can be effectively used to enlarge small training sets, such as DRIVE and CHASE\_DB1. Specifically, the AUC is improved by 0.17 and 0.34 on the DRIVE and CHASE\_DB1 datasets, respectively.
% è da rivedere su DRIVE ma sembrerebbe che il grosso vantaggio dei sintetici è nella sensitivity più alta...
\noindent A comparison with other state--of--the--art techniques applied to the two benchmarks, is reported in Table \ref{DRIVE_results} and \ref{CHASE_results}. 

\begin{table}[ht]
    \centering
    \fontsize{9}{11}\selectfont  
    \begin{tabular}{|l|l|l|}
    \hline
     Methods & AUC & Acc\\ \hline
    Jiang et al.\cite{25} & 96.80 \% & 95.93 \%\\ 
    Li et al. \cite{12} & 97.38 \% & 95.27 \%\\
    Dasgupta and Singh \cite{26} & 97.44 \% & 95.33 \%\\ 
    Yan et al. \cite{Yan2018JointSA} & 97.52 \% & 95.42 \%\\
    Mo and Zhang \cite{33} & 97.82 \% & 95.21 \%\\
    Liskowski and Krawiec \cite{11} & 97.90 \% & 95.35 \%\\
    Feng et al. \cite{27} & 97.92 \% & 95.60 \%\\
    Oliveira et al. \cite{OLIVEIRA2018229} & 98.21 \% & 95.76 \%\\
    Sekou et al. \cite{Sekou2019FromPT} & \textbf{98.74} \% & \textbf{96.90} \%\\
    Our & 98.65 \% & \textbf{96.90} \%\\
    \hline
    \end{tabular}
\vspace*{0.3cm}
\caption{\small{A comparison of vessel segmentation results on the DRIVE dataset.}}
\label{DRIVE_results}
\end{table}

\begin{table}[ht]
    \centering
    \fontsize{9}{11}\selectfont  
    \begin{tabular}{|l|l|l|}
    \hline
     Methods & AUC & Acc\\ \hline
    Jiang et al. \cite{25} & 95.80 \%  & 95.91 \%\\ 
    Li et al. \cite{12} & 97.16 \% & 95.81 \%\\
    Yan et al. \cite{Yan2018JointSA} & 97.81 \% & 96.10 \%\\
    Liskowski and Krawiec \cite{11} & 98.45 \% & 95.77 \%\\
    Mo and Zhang \cite{33} & 98.12 \% & 95.99 \%\\
    Oliveira et al. \cite{OLIVEIRA2018229} & 98.55 \% & 96.53 \%\\
    Sekou et al. \cite{Sekou2019FromPT} & 98.78 \% & 97.37 \%\\
    Our & \textbf{99.16} \% & \textbf{97.72} \%\\
    \hline
    \end{tabular}
\vspace*{0.3cm}
\caption{\small{A comparison of vessel segmentation results on the CHASE dataset.}}
\label{CHASE_results}
\end{table}

\noindent To compare the different retinal blood vessel segmentation methods is somehow difficult in datasets for which an explicit train--test split is not given (e.g CHASE\_DB1), because the split may differ from one paper to another. For instance, \cite{Yan2018JointSA} and \cite{12} use the same split employed in this paper, while \cite{33}, \cite{Sekou2019FromPT} and \cite{OLIVEIRA2018229} use a 4--fold cross--validation strategy (in \cite{OLIVEIRA2018229} each fold included 3 images of one eye and 4 images of the other), whereas \cite{11} only considers patches that are fully inside the field of view. However, our method demonstrates improved (or at least state--of--the--art) performances on both the DRIVE and CHASE\_DB1 datasets.

\section{Conclusions and future perspectives} \label{Conclusions}
In this paper, we have proposed a two stage procedure to generate synthetic retinal images. 
During the first stage, the semantic label masks, which correspond to the retinal vessels, are generated by a Progressively Growing GAN. Then an image--to--image translation approach is employed to obtain the retinal images from the label masks.
The proposed approach allows to generate images with unprecedented high resolution and
 realism.
The experiments demonstrate the usefulness of the synthetic images, that can be effectively used to train a deep segmentation network. Moreover, if a fine--tuning based on real images is applied after a preliminary learning phase based only on synthetic images, the performances of the segmentation network further improve, reaching or outperforming the state--of--the--art methods. 
It is worth noting that the proposed framework for image generation is general and not limited to retinal image generation.
It is a matter of further investigation the possibility of extending the proposed two--phase generation procedure to different domains.

%{\small
%\bibliographystyle{ieee}
%\bibliography{refs}
%}

\end{document}